\documentclass[twocolumn,showpacs,preprintnumbers,amsmath,amssymb]{revtex4}

\usepackage{graphicx}
\usepackage{dcolumn}
\usepackage{bm}
\usepackage{color}

\newtheorem{theorem}{Theorem}[section]\def\TH{\begin{theorem}}\def\HT{\end{theorem}}
\newtheorem{prop}[theorem]{Proposition}\def\PRO{\begin{prop}}\def\ORP{\end{prop}}
\newtheorem{coro}[theorem]{Corollary}\def\COR{\begin{coro}}\def\ROC{\end{coro}}
\def\COR{\begin{coro}}\def\ROC{\end{coro}}
\newtheorem{defi}[theorem]{Definition}\def\DE{\begin{defi}}\def\ED{\end{defi}}
\newtheorem{lemma}[theorem]{Lemma}\def\LE{\begin{lemma}}\def\EL{\end{lemma}}
\newtheorem{algo}{Algorithm}\def\AL{\begin{algo}}\def\LA{\end{algo}}
\newtheorem{example}{Example}\def\EX{\begin{example}}\def\XE{\end{example}}

\def\ket#1{{|}#1\rangle}
\def\bra#1{\langle#1{|}}

\definecolor{Ocolor}{rgb}{.9,.1,.1}
\definecolor{2color}{rgb}{.2,.5,.2}
\definecolor{3color}{rgb}{0,0,1}

\def\elham#1{\textcolor{2color}{#1}}

\begin{document}

\title{Closed timelike curves in measurement-based quantum computation}

\author{Raphael Dias da Silva, Ernesto F. Galv\~{a}o}
 \affiliation{Instituto de F{\'i}sica, Universidade Federal
Fluminense \\Av. Gal. Milton Tavares de Souza s/n\\Gragoat{\'a},
Niter{\'o}i, RJ, 24210-340, Brazil}
\author{Elham Kashefi}%
\affiliation{%
School of Informatics, University of Edinburgh, Edinburgh EH8 9AB, U.K.
}

\date{\today}

\begin{abstract}

Many results have been recently obtained regarding the power of hypothetical closed time-like curves (CTCs) in quantum computation. Here we show that the one-way model of measurement-based quantum computation 
encompasses in a natural way the CTC model proposed by Bennett, Schumacher and Svetlichny. We identify a class of CTCs in this model that can be simulated deterministically and point to a fundamental limitation of Deutsch's CTC model 
 which leads to predictions conflicting with those of the one-way model.

\end{abstract}
\maketitle

\section{Introduction}
The possibility of time travel has been studied for decades in the context of general relativity. Assuming that closed time-like curves (CTCs) exist, a series of results were obtained regarding their implications for quantum mechanics and quantum computation \cite{Deutsch91, Bacon03, BrunHW09, BennettLSS09, Ralph07, RalphM10}.

In this paper we describe how the one-way model of measurement-based quantum computation \cite{RB01} encompasses in a natural way a model for CTCs proposed by Bennett and Schumacher \cite{BennettS02}, and more recently by Svetlichny \cite{Svetlichny09}. We show that the one-way model effectively simulates deterministically a class of CTCs in this model, and characterize this class. A second model for CTCs is Deutsch's highly influential study of quantum time-travel \cite{Deutsch91}.  We show that Deutsch's model leads to predictions conflicting with those of the one-way model, and identify the reason behind this.

The paper is organized as follows. In section \ref{sec bss} we review the quantum CTC model based on teleportation and postselection proposed by Bennett, Schumacher and Svetlichny. In section \ref{sec ctcmbqc} we discuss how CTCs appear naturally in the one-way model of measurement-based quantum computation, and show that they correspond to CTCs in the Bennett/Schumacher/Svetlichny model. In section \ref{sec deutsch} we review the quantum CTC model due to Deutsch, and contrast it with the CTCs that appear in the one-way model.

\section{A model for CTCs based on teleportation and post-selection}\label{sec bss}

In this section we briefly review the main features of the CTC model proposed by Bennett and Schumacher \cite{BennettS02} and by Svetlichny \cite{Svetlichny09}, referred to from now on as the Bennett/Schumacher/Svetlichny (BSS) model. Ideas similar to the BSS model were proposed independently also by Horowitz and Maldacena \cite{HorowitzM04} in the context of black hole evaporation \cite{GottesmanP04}. In the recent preprint \cite{Lloydetal10} there appeared a discussion of some characteristics of this model, together with experimental simulation of a particular CTC.

\begin{figure} 
 \centering
\includegraphics[scale=.4]{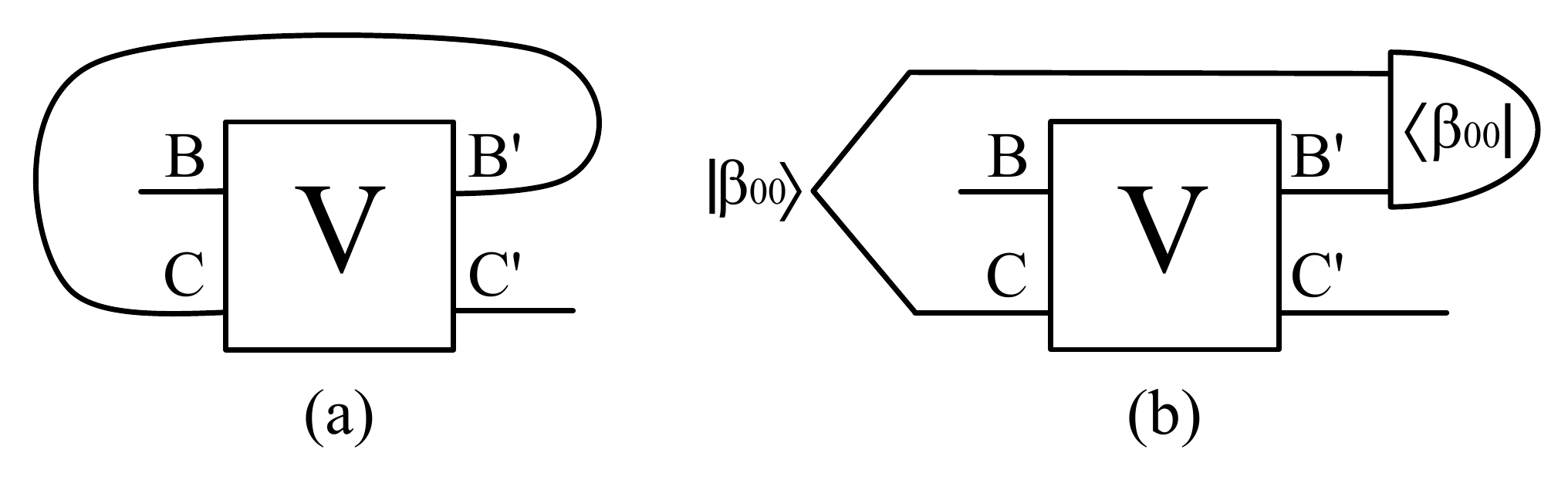}
 \caption{a) CTC takes qubit back in time to interact with its past self; b) Bennett/Schumacher/Svetlichny (BSS) circuit to simulate this CTC probabilistically, using teleportation and post-selection (see main text).}
 \label{fig bsstravel}
\end{figure}

For simplicity, we will restrict our discussion to two-qubit unitaries; the generalization to larger-dimensional systems is straightforward. In Fig. \ref{fig bsstravel}-a we represent a circuit with a CTC that takes the top qubit back in time to interact with its past self via the two-qubit unitary $V$. This CTC is simulated using teleportation in BSS's construction (see Fig. \ref{fig bsstravel}-b). Two qubits are prepared in the Bell state $\ket{\beta_{00}}=1/\sqrt{2}(\ket{00}+\ket{11})$, with one of them sent though $V$ together with an arbitrary input state $\ket{\psi_{in}}$ at position B in Fig. \ref{fig bsstravel}-b. After $V$ we perform a Bell-state measurement, postselecting those events corresponding to projection onto the initial $\ket{\beta_{00}}$. The post-selected teleportation guarantees that the state at $C$ is state $B'$ teleported back in time to interact with state $B$ via $V$. The scheme works only probabilistically, implementing a map from state $B \to C'$.

In the absence of yet-undiscovered physical CTCs, quantum circuits such as the BSS circuit in Fig. \ref{fig bsstravel}-b simulate the CTCs with a finite probability of success.  Svetlichny's model differs from that of Bennett and Schumacher in an irrelevant detail only: the unitaries he considered involved a swap between the states to be fed to $V$, that is, he modelled CTCs such as the one in Fig. \ref{fig bsstravel}-a with unitaries of the form $V=U\cdot SWAP$.

\begin{figure}
 \centering
\includegraphics[scale=.4]{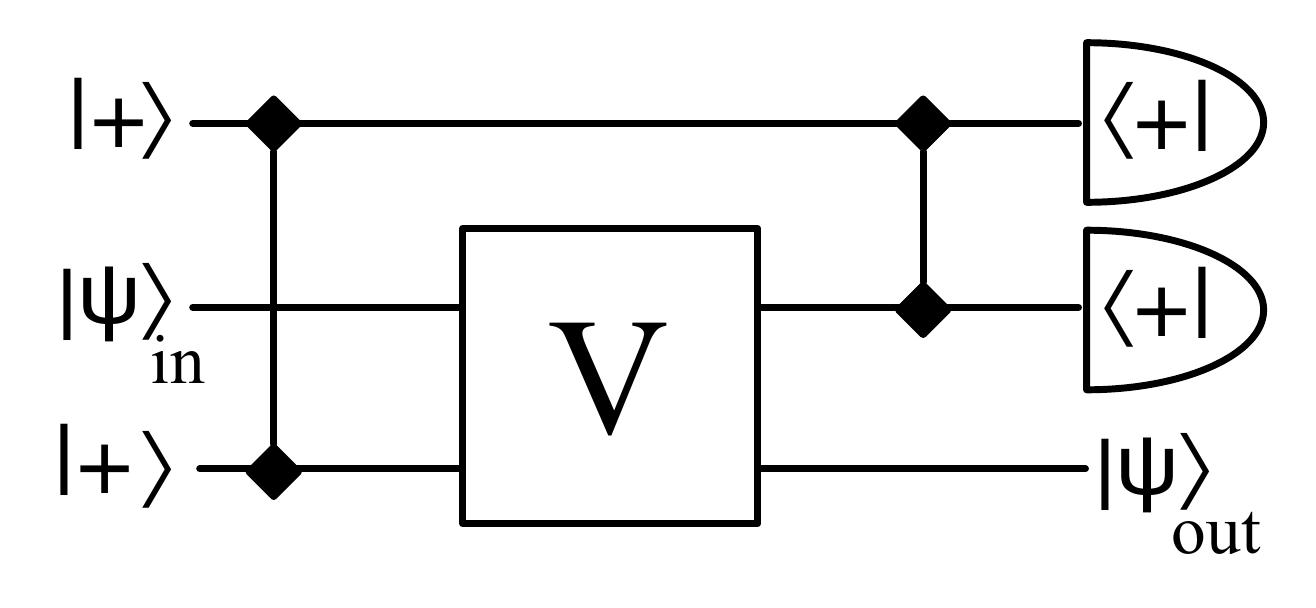}
 \caption{Rewritten BSS circuit using preparation of and projections onto $\ket{+}$ states. The unitary $V$ is decomposed using the universal gate-set consisting of $J_{\theta}$ and ${\wedge}Z$.}
 \label{fig bssequiv1}
\end{figure}

In order to link the BSS CTC simulation circuits with the one-way model of quantum computation \cite{RB01} we consider the universal gate set defined with single-qubit gate $J_{\theta} \equiv \frac{1}{\sqrt{2}}\left(
\begin{array}{cc}
1 & e^{i\theta}\\
1 & -e^{i\theta}
\end{array}
\right)
$ and controlled-$Z$ gate ${\wedge}Z \equiv 1-2\ket{11}\bra{11}$ \cite{DanosKP05}. We can rewrite any BSS circuit using these gates, initialization in state $\ket{+}\equiv 1/\sqrt{2}(\ket{0}+\ket{1})$, and final Pauli $X$ measurements that postselect projections onto state $\ket +$ (see Fig. \ref{fig bssequiv1}). This is closely related to the setting of the one-way model, so we can use various tools developed in that context to study CTCs as described by the BSS model.

\begin{figure}
 \centering
\includegraphics[scale=.4]{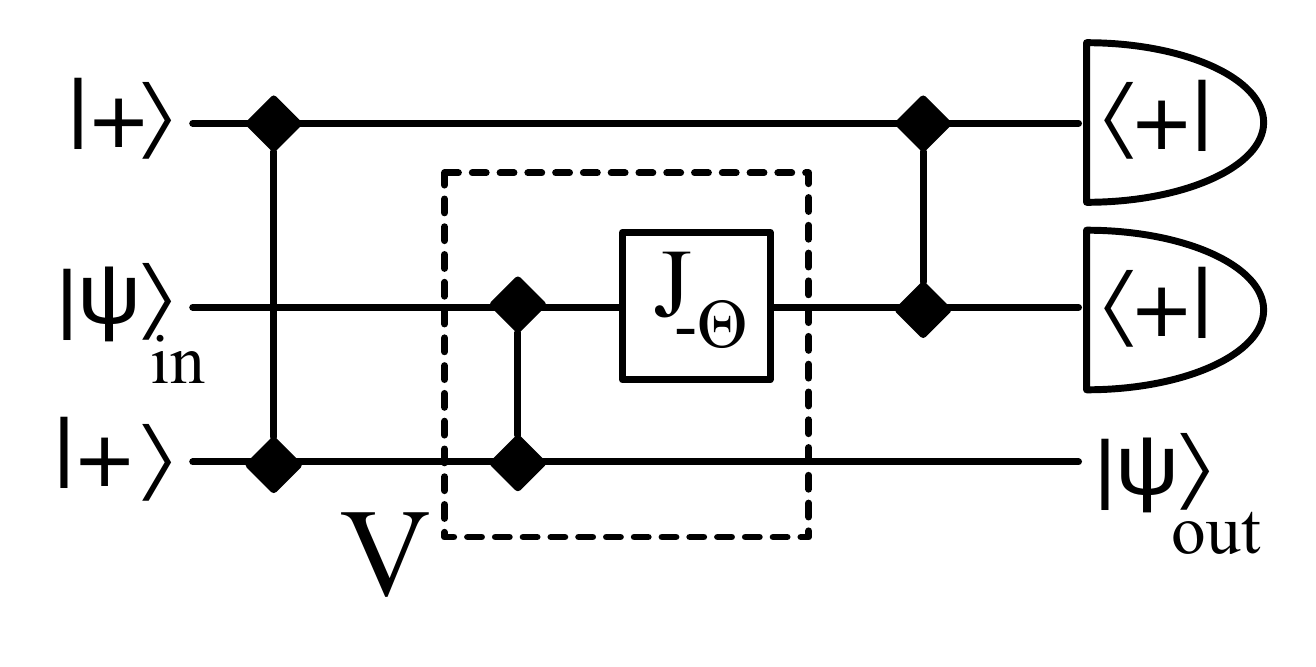}
 \caption{BSS circuit simulating a CTC with unitary $V=(J_{-\theta} \otimes 1) {\wedge}Z$.}
 \label{fig bssplus}
\end{figure}

Let us work out the BSS simulation for one particular CTC of interest, given by unitary $V=(J_{-\theta} \otimes I) {\wedge}Z$. The BSS circuit that simulates this particular CTC is shown in Fig. \ref{fig bssplus}. The circuit acts on input state $\ket{\psi_{in}}=\alpha\ket{0}+\beta\ket{1}$ to output state $\ket{+}$ with probability $|\alpha+e^{-i\theta}\beta|^2/4$, as can be easily checked. This means the BSS formalism predicts that the action of this CTC is to deterministically project the input state onto $\ket{+}$.

Interestingly, the finite probability of success is the mechanism that avoids the grandfather paradox, i.e. situations in which the combination of input state and interaction $V$ prevents the existence of a self-consistent state for the time-travelling system. In such situations, the BSS model yields a probability of success equal to zero, as noted in \cite{BennettS02, Lloydetal10}.

\section{CTCs in the one-way model} \label{sec ctcmbqc}

In this section we show how CTCs appear naturally in the one-way model of quantum computation \cite{RaussendorfBB01}. The key element is the appearance of anachronical dependencies, which in previous works \cite{DanosK06,BrowneKMP07} were dealt with formally to obtain physically doable operations corresponding to the implementation of a deterministic computation. Here we study these problematic time dependencies in more detail, and show that they correspond to CTCs as modelled by the Bennett/Schumacher/Svetlichny (BSS) model discussed in the last section.

In what follows we will describe measurement-based computations using the formal language known as \textit{Measurement Calculus} \cite{DanosKP07}, using a simplified version of it which will be enough for our purposes here. Let $M_i^{\theta}$ represent a measurement on qubit $i$ onto basis $\{ \ket{\pm_{\theta}}\equiv 1/\sqrt{2}(\ket{0} \pm e^{i\theta}\ket{1})\}$, with outcome $s_i=0$ associated with $\ket{+_{\theta}}$, and $s_i=1$ with $\ket{-_{\theta}}$. $X_i^{s_j}$ represents a Pauli $X$ operator acting on qubit $i$, controlled by the classical outcome of the measurement on qubit $j$, and similarly for $Z_i^{s_j}$. Finally, the operator $N_i$ represents initialization of qubit $i$ in state $\ket{+}$, which can be entangled with other qubits with the controlled-$Z$ gate $\wedge Z$. These operations can be put together as time-ordered sequences of operations \elham(called \emph{measurement patterns}), which form words of the formal language. The meaning (semantics) of a pattern is the map that it implements between input and output qubits. Of course some patterns correspond to unphysical operations, and only well-formed-formulas representing physically realizable operations have meaning.

\begin{figure}
 \centering
 \includegraphics[scale=0.4]{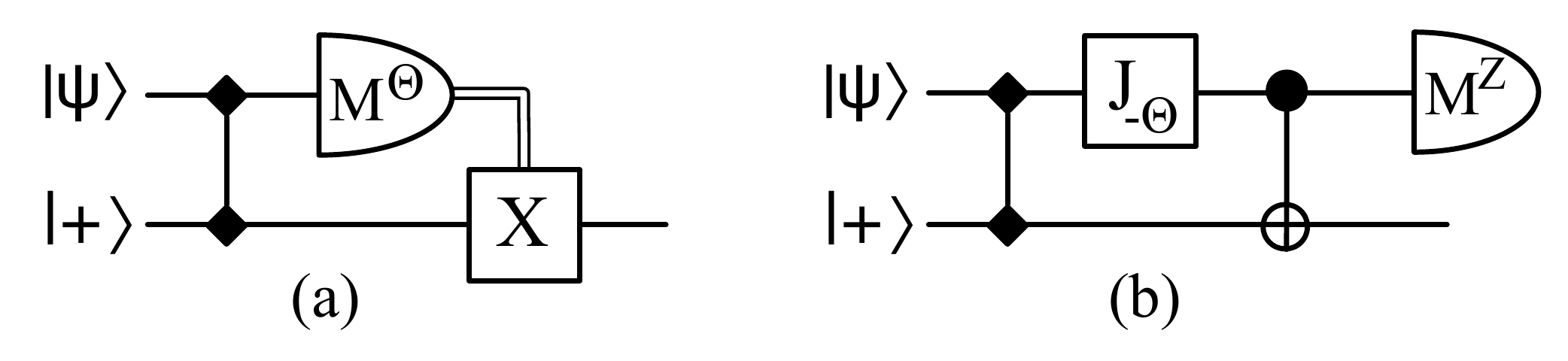}
 \caption{Two equivalent circuits, i.e. implementing the same unitary. In a) we have a classically-controlled $X$ unitary dependent on the measurement outcome of $\ket{\pm_{\theta}}$ basis projection. In b) this has been turned into a coherent circuit with measurement onto the $Z$ basis.}
 \label{fig jcc}
\end{figure}

For concreteness, let us start by analyzing a simple pattern of operations implementing a one-qubit unitary: 
\begin{equation}
X_2^{s_1}M_1^{\theta}{\wedge}Z_{12} N_2 \ket{\psi_{in}}_1\label{seq1}
\end{equation}
This sequence of operations can be represented as a two-qubit quantum circuit, see Fig. \ref{fig jcc}-a. An arbitrary input state $\ket{\psi_{in}}_1$, previously entangled via a ${\wedge}Z$ gate with a qubit initially in state $\ket{+}_2$, is then measured in the $\ket{\pm_{\theta}}$ basis. The outcome $s_1=0$ or $1$ controls classically whether or not to apply a Pauli $X$ gate on qubit 2. Fig. \ref{fig jcc}-b represents the same operations, only with a $Z$ basis measurement and with the controlled operation implemented coherently as a ${\wedge}X$ (CNOT) gate. The two circuits are equivalent as they implement the same unitary $J_{-\theta}$ on initial state $\ket{\psi_{in}}_1$, with output in qubit 2.

\begin{figure}
 \includegraphics[scale=0.4]{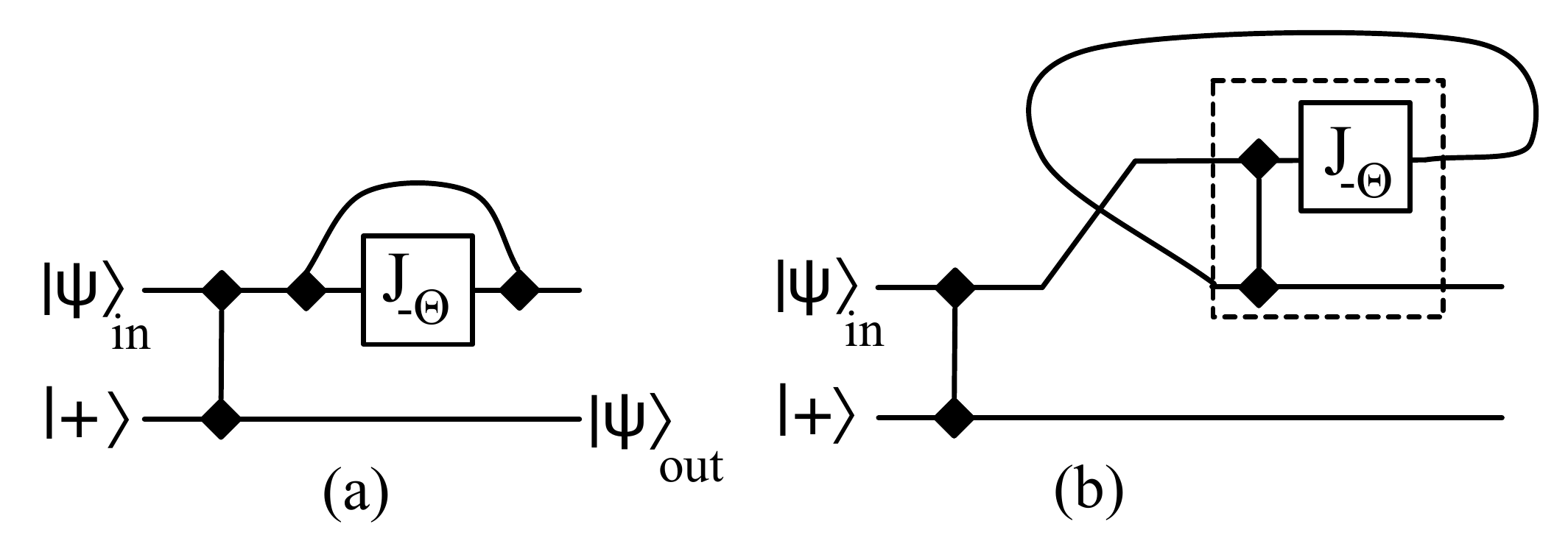}
 \caption{a) Circuit that includes a CTC with an anachronical ${\wedge}Z$ gate; it is equivalent to the two circuits in Fig. \ref{fig jcc}. b) 
 The same circuit rewritten in the BSS format.}
 \label{fig jloop2}
\end{figure}

It is easy to find other patterns (and corresponding circuits) which are equivalent to sequence (\ref{seq1}), i.e. implement the same unitary $J_{-\theta}$ between input and output qubits. We start by observing that the state to be measured $\ket{G}={\wedge}Z_{12}\ket{\psi_{in}}_1\ket{+}_2$ is stabilized by the two operators $\{ 1=Z_1^0X_2^0,Z_1^1X_2^1 \}$, as $Z_1X_2\ket{G}=\ket{G}$. In other words, $Z_1^{s_1}X_2^{s_1}$ is a stabilizer of $\ket{G}$ independently of whether $s_1=0$ or $1$. This enables us to manipulate sequence (\ref{seq1}) as follows:
\begin{equation}
X_2^{s_1} M_1^{\theta} \ket{G} = X_2^{s_1} M_1^{\theta} Z_1^{s_1} X_2^{s_1} \ket{G} = M_1^{\theta} Z_1^{s_1} \ket{G}. \label{eq stab1}
\end{equation}
This last sequence represents a time-travel conundrum: a classically controlled Pauli $Z$ unitary which must be applied depending on the outcome of an as-yet unmeasured qubit. This is turned into a quantum CTC if we apply the anachronical Pauli $Z$ operation coherently,  as we see in Fig. \ref{fig jloop2}-a. A rewriting of this circuit in slightly different form (Fig. \ref{fig jloop2}-b) shows that the top qubit enters exactly the CTC we analyzed using the BSS model (see Fig. \ref{fig bssplus}). We can now compare the predictions of the one-way model for this particular CTC with those given by the BSS CTC model.

An apparent mismatch between the BSS formalism and the one-way model appears when we analyze the action of the CTC in Fig. \ref{fig jloop2}-b. On the one hand, our analysis of the circuit in Fig. \ref{fig bssplus} has shown that the CTCs effect is to project the input state onto $\ket{+}$. On the other hand, comparison between Figs. \ref{fig jcc}-a and \ref{fig jloop2}-a suggests that the effect of the CTC should be to project the top qubit onto $\ket{+_{\theta}}$ instead. This is because a post-selected $\ket{+_{\theta}}_1$ projection in the circuit of Fig. \ref{fig jcc}-a is what it takes to implement the unitary $J_{-\theta}$ to $\ket{\psi_{in}}$, without the need for the controlled $X$ correction.

\begin{figure}
\includegraphics[scale=0.4]{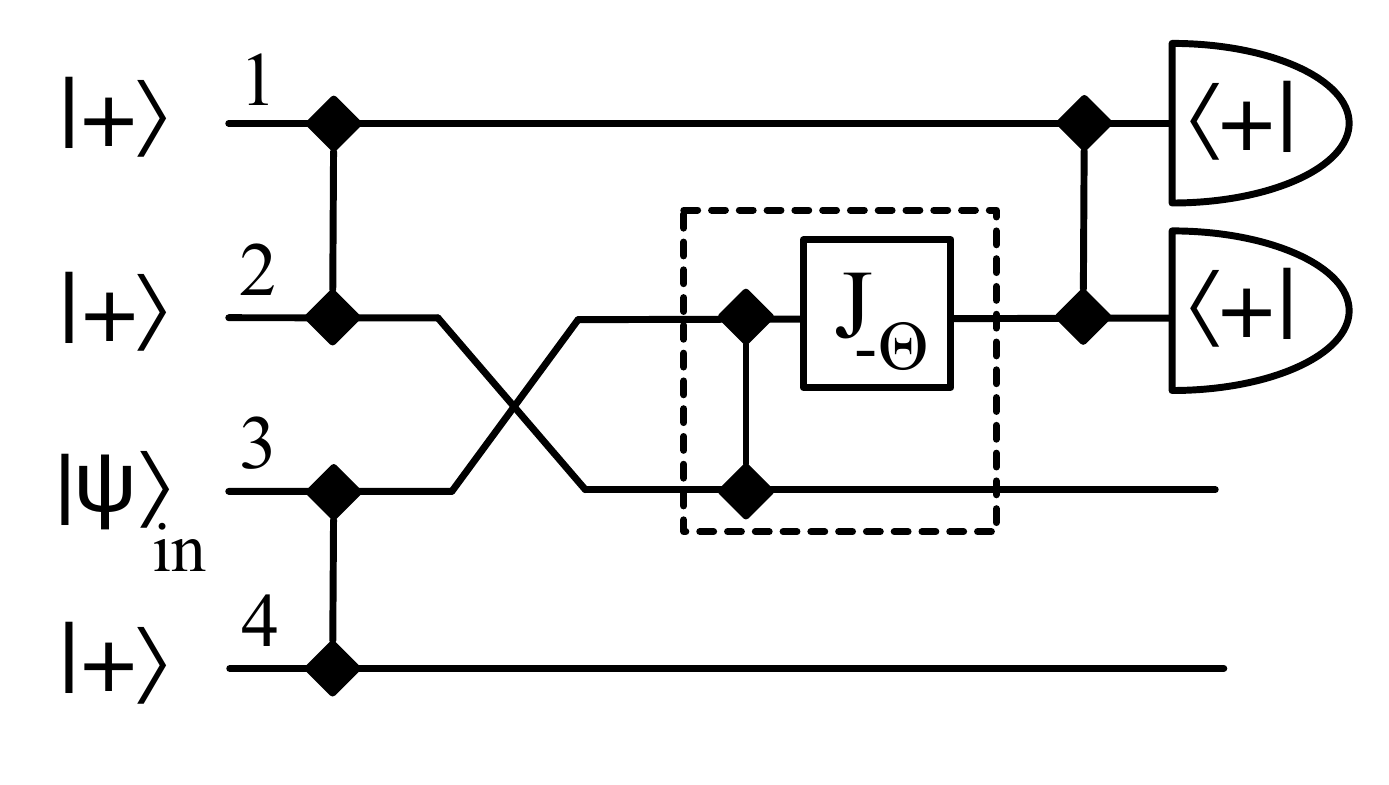}
\caption{Circuit that implements the probabilistic BSS simulation of the CTC circuit in Fig. \ref{fig jloop2}-b.}
 \label{fig bigbss}
\end{figure}

The resolution of this apparent conflict is surprising. The one-way model only predicts that, when embedded in the circuit of Fig. \ref{fig jloop2}-a, the CTC should implement the same input-output map $J_{-\theta}$ as its two equivalent circuits in Fig. \ref{fig jcc}. Using the circuit in Fig. \ref{fig bssplus} to simulate what happens in the CTC of Fig. \ref{fig jloop2} is unwarranted; instead, we should simulate the CTC's action when embedded in the circuit of Fig. \ref{fig jloop2}. The BSS circuit for this simulation is in Fig. \ref{fig bigbss}. A simple calculation shows that this circuit, at once, fulfills the predictions of both the BSS and the one-way model: it projects qubits $3$ and $4$ onto state $\ket{+}_3\otimes (J_{-\theta}\ket{\psi_{in}})_4$.

This illustrates what seems to be a general feature of CTCs: their effect extends not only to the time-travelling sub-system $A$, but to all sub-systems that have interacted with $A$ prior to $A$'s encounter with CTCs. In the context of the one-way model what interests us is the dynamics the measured (time-travelling) qubits induce on the output (time-respecting) qubits. It is this dynamical map that we can calculate and compare, as we have done in this section. This will also be the key that allows for the comparison with Deutsch's CTC model in section \ref{sec deutsch}.

\subsection{A second CTC example}

\begin{figure}
 \includegraphics[scale=0.16]{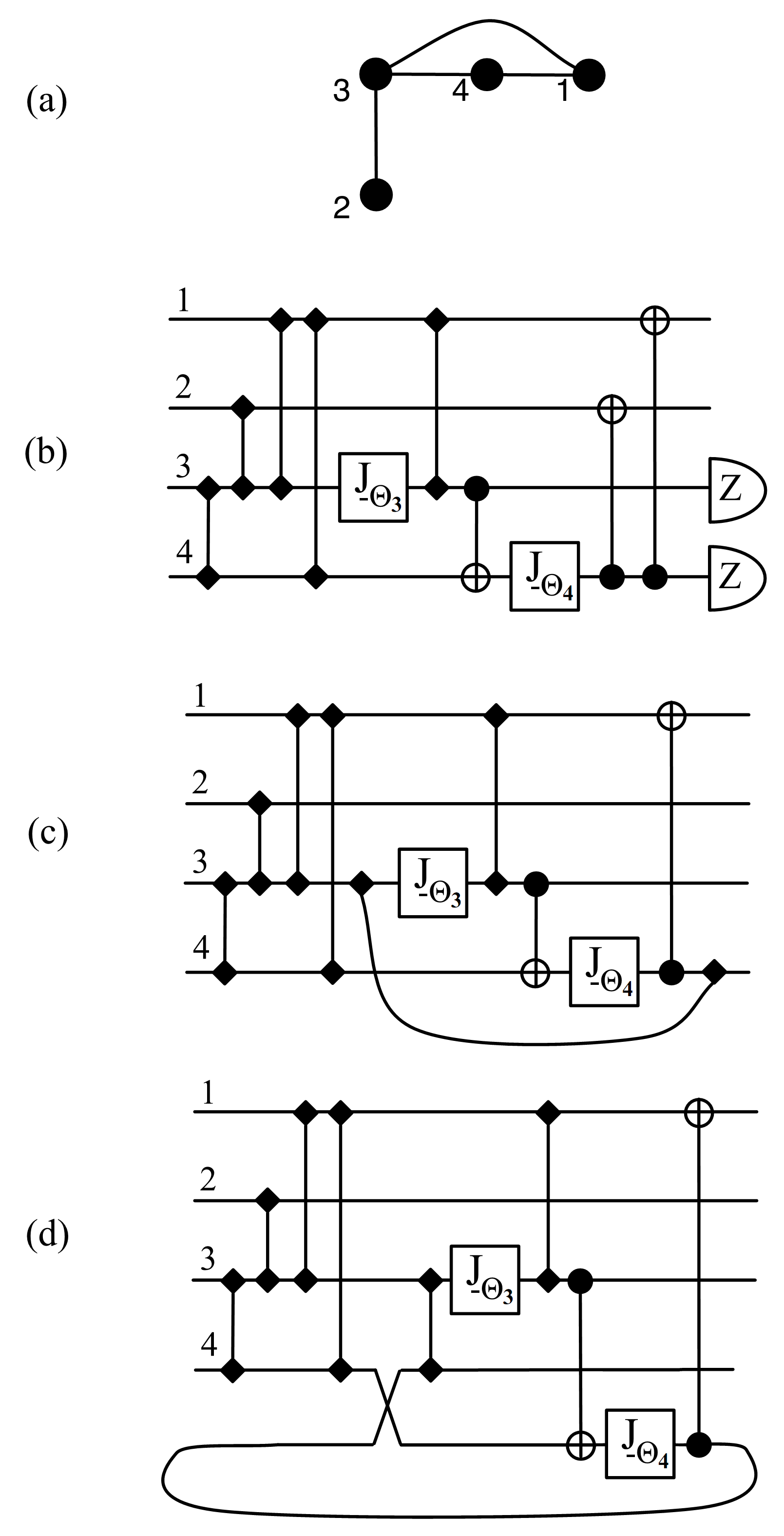}
 \caption{a) Entanglement graph corresponding to state $\ket{G}$ in eq. (\ref{eq g2}). On this state we can perform the sequence of one-way operations in eq. (\ref{eq example}). b) Straightforward circuit translation of the sequence of operations in eq. (\ref{eq example}). c) CTC circuit corresponding to the same computation, obtained by using $\ket{G}=K_2 \ket{G}$, with $K_2$ given by eq. (\ref{eq k2}). d) Redrawing of the circuit in c) to explicitly show the CTC that takes qubit 4 back in time.} \label{fig ctc2}
\end{figure}

The simple form of the one-way pattern in eq. (\ref{seq1}) may suggest that the CTCs that appear in the one-way model are only as simple as the one appearing in Fig. \ref{fig jloop2}, with CTC unitary $V$ consisting of only two gates, a ${\wedge}Z$ and a $J_{\theta}$. In fact, a given deterministic one-way computation can be equivalent to the simulation of different CTCs implementing the same input-output map, and these CTCs may have different structures. To illustrate this,  let us consider the following sequence of commands implementing a one-way computation:
\begin{equation} \label{eq example}
 X_2^{s_4} X_1^{s_4} Z_1^{s_3} M_4^{\theta_4} X_4^{s_3} M_3^{\theta_3} \ket{G} ,
\end{equation}
where
\begin{equation} \label{eq g2}
\ket{G} = {\wedge}Z_{23} {\wedge}Z_{13} {\wedge}Z_{14} {\wedge}Z_{34} N_2 N_1 \ket{\psi_{in}}_{34}
\end{equation}
is the state associated with the graph in Fig. \ref{fig ctc2}-a. As before, $M_i^{\theta}$ represents a measurement on qubit $i$ onto basis $\{ \ket{\pm_{\theta}}\equiv 1/\sqrt{2}(\ket{0} \pm e^{i\theta}\ket{1})\}$, with outcome $s_i=0$ associated with $\ket{+_{\theta}}$, and $s_i=1$ with $\ket{-_{\theta}}$. In this graph, vertices represent qubits and edges represent ${\wedge}Z$ interactions between them, which create the entanglement structure that is exploited by the one-way quantum computation.  A step-by-step description of how to obtain a deterministic one-way pattern for a given entanglement graph can be found in \cite{BrowneKMP07}.

We can do with this sequence of commands what we did with the sequence in (\ref{seq1}), translating each measurement as a $J_{\theta}$ gate followed by $Z$ projections, and including corrections as coherent ${\wedge}Z$ and ${\wedge}X$ gates. This results in a straightforward translation of the sequence of operations into a circuit, which we show in Fig. \ref{fig ctc2}-b. Note that as the sequence in (\ref{eq example}) is time-respecting, so is the associated circuit.

We can now obtain different sequences of one-way operations that implement the same map, by rewriting the initial state $\ket{G}=K_i \ket{G}$, with $K_i$ being a stabilizer of $\ket{G}$. It is easy to check that the following operators are stabilizers of $\ket{G}$: 
\begin{eqnarray} \label{stabilizer}
 K_1 &=& X_1^{s_4} Z_3^{s_4} Z_4^{s_4} \\
 K_2 &=& X_2^{s_4} Z_3^{s_4} \label{eq k2}\\ 
 K_4 &=& X_4^{s_3} Z_3^{s_3} Z_1^{s_3}
\end{eqnarray}
Note that as qubit 3 is in an arbitrary input state, $K_3=X_3 Z_1 Z_2 Z_4$ is not a stabilizer of $\ket{G}$. Using stabilizers $K_1, K_2, K_4$ we obtain three new sequences of operations that now include anachronical corrections. For example, after applying $K_2^{s_4}$ we have the following pattern:
\begin{eqnarray} \nonumber
&&X_2^{s_4} X_1^{s_4} Z_1^{s_3} M_4^{\theta_4} X_4^{s_3} M_3^{\theta_3} \; X_2^{s_4} Z_3^{s_4} \; \ket{G} \\ \nonumber
&=&  X_1^{s_4} Z_1^{s_3} M_4^{\theta_4} X_4^{s_3} M_3^{\theta_3} Z_3^{s_4} \ket{G}.
\end{eqnarray}
Each such sequence can be translated into the circuit model, where the classically controlled $X$ and $Z$ corrections appear as coherent ${\wedge}X$ and ${\wedge}Z$ gates. These anachronical gates correspond to CTCs, as we illustrate for the case of $K_2$ in Fig. \ref{fig ctc2}-c. In Fig. \ref{fig ctc2}-d we redraw the CTC circuit in \ref{fig ctc2}-c so as to explicitly show the CTC that takes qubit 4 back in time.

Any deterministic one-way pattern yields a class of CTC circuits simulatable by it. These circuits are obtained as shown above, by using arbitrary stabilizers that introduce the anachronical dependencies in the (originally deterministic and time-respecting) pattern.

\subsection{Deterministic simulations of CTCs}

Note that a deterministic simulation of the BSS circuit in Fig. \ref{fig bigbss} (with respect to its action on input state $\ket{\psi_{in}}$) is achieved by the circuit in Fig. \ref{fig jcc}-b, which effectively implements unitary $J_{-\theta}$. In other words, the circuit in Fig. \ref{fig jcc}-b simulates deterministically the CTC circuit in Fig. \ref{fig jloop2}-b. A natural question is then to determine which BSS CTCs can be simulated deterministically by a one-way pattern and its equivalent circuit. We now present a systematic way to find measurement patterns that deterministically simulate CTCs in BSS's model.

We start with the BSS circuit in which each CTC is simulated by preparation of state $\ket{\beta_{00}}$ and subsequent postselected projection onto the same state. The next step is to translate the BSS circuit into a one-way measurement pattern, which can always be done using the well-known techniques described in \cite{BroadbentK09}. The main difference between the patterns in \cite{BroadbentK09} and ours concerns the translation of the post-selected Bell-pair measurements, which are translated as post-selected Pauli $X$ measurements. Moreover we translate any deterministic $J$ gate as a deterministic projection, represented by a measurement with an anachronical correction (as justified in \cite{BroadbentK09}). 

The resulting pattern implements the same input-output map as the original BSS circuit and it can be manipulated using stabilizer operations (such as local complementation \cite{vandenNestDM04}), with the aim of eliminating the ancillas added in the $\ket{\beta_{00}}$ state preparations and postselections required by the BSS simulation circuit. This can always be done since any such Bell-pair projections translate only as a sequence of Pauli projections, which enables us to apply the general rules for removing a Pauli measurement from a measurement pattern \cite{HeinDERvdNB06}. In some cases this results in a pattern where the anachronical corrections (added during the translation of the $J$ gate) can no longer be eliminated, resulting in anachronical circuits corresponding to unsound physical operations.

In other cases, however, the resulting pattern satisfies the determinism conditions for the one-way model obtained in \cite{DanosK06, BrowneKMP07}, where they were called \emph{flow} and \emph{generalised-flow}, respectively. If that is the case, the anachronical corrections that appear can be removed. Note that the operations used in removing the ancillas introduced by the BSS simulations consist of Local Complementation, removal of Pauli measurements, and other stabilizer manipulations, all of which preserve the map implemented from input to output. As a result, the newly-found sequence (and its equivalent circuit) implements deterministically the same map that succeeded only probabilistically in the BSS simulation circuit. This effectively characterizes a class of BSS CTC circuits that admit a deterministic simulation in the one-way model.

As an example, in the Appendix we work out explicitly the stabilizer manipulations required to obtain the deterministic simulation of the BSS circuit in Fig. \ref{fig bigbss}, and give more details on the general translation scheme.

\section{Conflict with Deutsch's CTC model}\label{sec deutsch}

In 1991 Deutsch \cite{Deutsch91} proposed a different, highly influential model for CTCs in quantum theory, which we now turn to. Deutsch's model avoids paradoxes by demanding self-consistent solutions for the time-travelling systems. Let us recall the main features of this model using just one time-respecting qubit and one time-travelling qubit, illustrated in Fig. \ref{fig deutsch}-a, where $U$ is a general two-qubit unitary. The correspondence with BSS's model is shown in Fig. \ref{fig deutsch}-b, so that Deutsch's $U=V\cdot SWAP$, with $V$ being the unitary in BSS's formulation of CTCs (Fig. \ref{fig bsstravel}).

No paradox arises if we demand that the time-travelling qubit state $\rho_{CTC}$ be a fixed point of the dynamics:
\begin{figure}
 \centering
 \includegraphics[scale=.40]{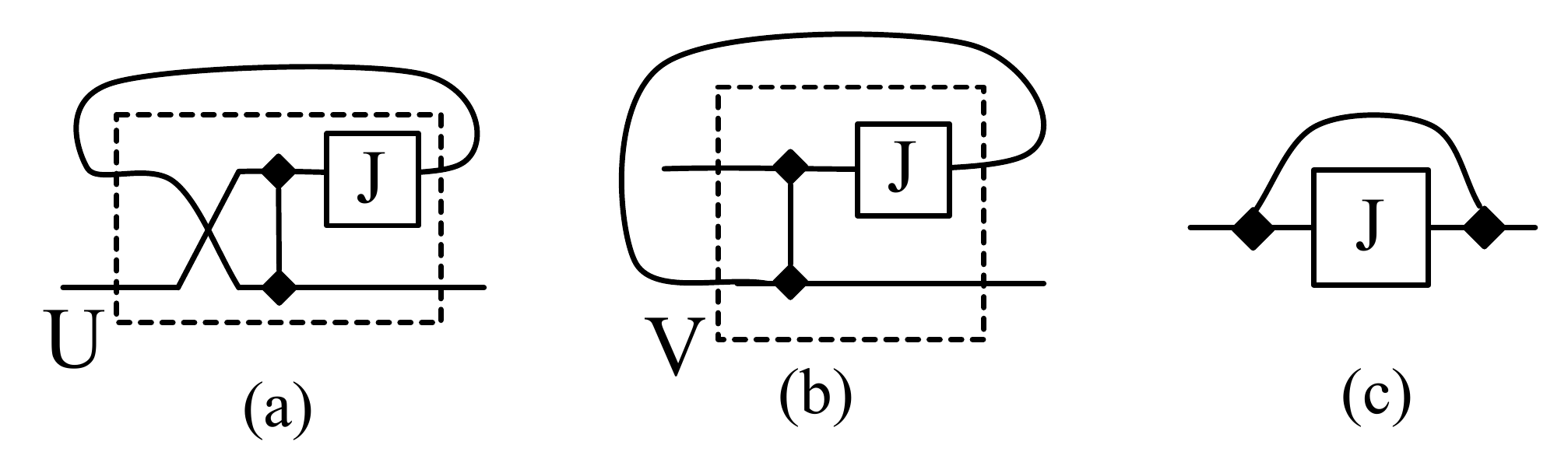}
 \caption{a) Deutsch's model for a CTC. b) This is the relationship between Deutsch's unitary $U$ and unitary $V$ in the BSS CTC circuit of Fig. \ref{fig bsstravel}-b.}
 \label{fig deutsch}
\end{figure}
\begin{equation}
\rho_{CTC}=Tr_{TR}\left( U (\rho_{CTC}\otimes \rho_{in}) U^{\dagger}\right), \label{eq sc}
\end{equation}
where the partial trace is over the time-respecting qubit. This self-consistency requirement defines multiple solutions for $\rho_{CTC}$, each of which corresponds to a (generally non-linear) map on $\rho_{in}$, which can be worked out from the solution $\rho_{CTC}$.

Let us now study the same CTC we analyzed using BSS's model, but now using Deutsch's by setting $U=(J_{-\theta} \otimes 1)\cdot {\wedge}Z \cdot SWAP$ in Fig. \ref{fig deutsch}.  In the BSS model this corresponded to the circuit in Fig. \ref{fig bssplus}. Three graphical representations for the same CTC are shown in Fig. \ref{fig 3ctcs}. We can represent $\rho_{in}(\vec{n})=1/2(1+\vec{n} \cdot \vec{\sigma})$ and $\rho_{CTC}(\vec{m})=1/2(1+\vec{m} \cdot \vec{\sigma})$, using the Pauli matrices $\vec{\sigma}=(X, Y,Z)$. A simple calculation using eq. (\ref{eq sc}) gives us the consistency conditions:
\begin{figure}
 \includegraphics[scale=0.38]{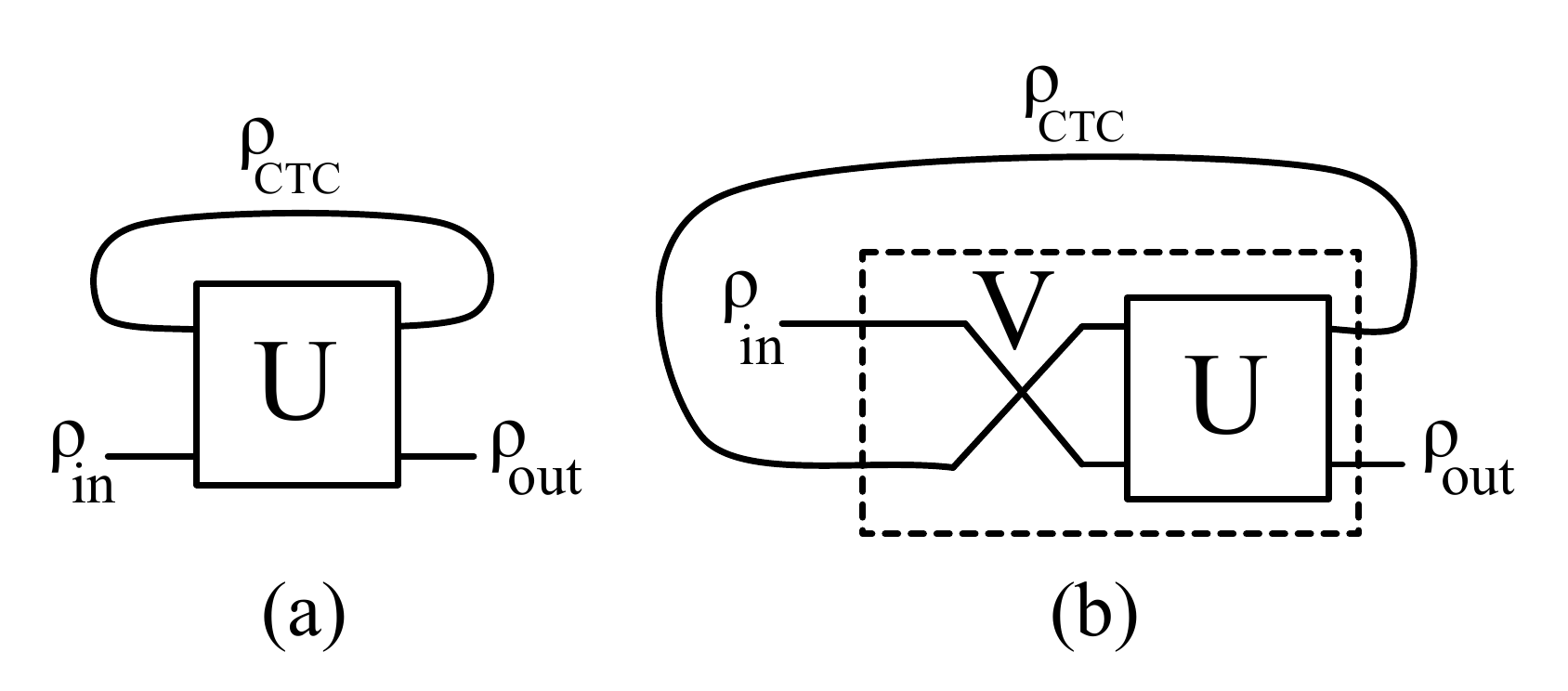}
 \caption{Three representations for the same CTC circuit. a) Deutsch formulation. b) BSS formulation. c) Short-hand form of either.}
 \label{fig 3ctcs}
\end{figure}
\begin{eqnarray}
m_x&=& n_z, \label{eq consist1}\\
m_y&=& m_z(n_x \sin\theta-n_y \cos\theta), \label{eq consist2} \\
m_z&=& m_z(n_x \cos\theta+n_y \sin\theta). \label{eq consist3}
\end{eqnarray}
The output state is $\rho_{out}(\vec{r})=1/2(1+\vec{r} \cdot \vec{\sigma})$, with $\vec{r}=(m_x n_z, m_y n_z, m_z)$ being a function of self-consistently assigned $m_x, m_y, m_z$.

There are two classes of self-consistent solutions to eqs. (\ref{eq consist1})-(\ref{eq consist3}). The first is obtained by setting $m_z=0$, which yields a unique self-consistent $\rho_{CTC}$ for each input state:
\begin{eqnarray}
\rho_{CTC}&:& \vec{m}=(n_z,0,0), \label{eq actc}\\
\rho_{out}&:& \vec{r}=(n_z^2,0,0).\label{eq aout} 
\end{eqnarray}
These solutions are valid for all input states $\rho_{in}(\vec{n})$. The second class of 
solutions is obtained by assuming that $m_z \neq 0$ in eqs. (\ref{eq consist1})-(\ref{eq consist3}). Self-consistency dictates that such solutions exist only for the particular $\rho_{in}= \ket{+_{\theta}}\bra{+_{\theta}}$, with $\rho_{CTC} = \rho_{out}$ described by $\vec{m}=(0,0,m_z)$.

\begin{figure}
\includegraphics[scale=0.4]{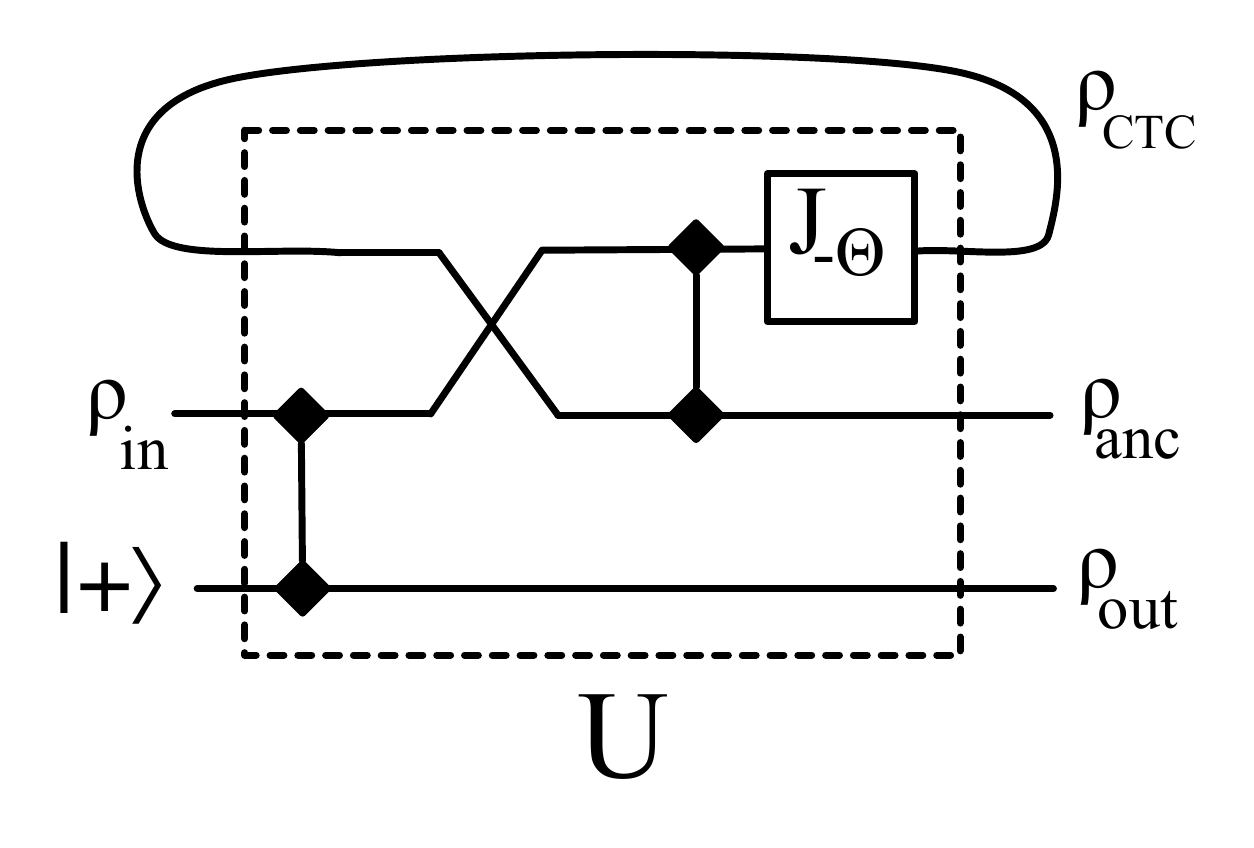}
\caption{Deutsch's formulation of the extended CTC circuit of Fig. \ref{fig jloop2}-b.}
\label{fig bigdeutsch}
\end{figure}

Our analysis of the BSS model for this CTC considered not only the circuit where the CTC appears on its own (Fig. \ref{fig 3ctcs}), but also an enlarged circuit where the CTC acts on only part of a larger entangled state (Fig. \ref{fig jloop2}). For a fair comparison between the two models, this can be done also in the Deutsch model, see Fig. \ref{fig bigdeutsch}. In this second approach Deutsch's unitary $U$ is a three-qubit unitary encompassing all gates in the circuit of Fig. \ref{fig jloop2}-b. A straighforward calculation using the self-consistency conditions (\ref{eq sc}) yields the solution in which $\rho_{CTC}$ and $\rho_{out}$ are parametrized by $\vec{m}=(n_z,0,0)$, with the ancilla qubit in output state $\rho_{anc}(\vec{a}): \vec{a}=(n_z^2,0,0)$.

Our results for both the smaller circuit of Fig. \ref{fig 3ctcs}-a and the larger circuit in Fig. \ref{fig bigdeutsch} show that Deutsch's model fails to implement the same input-output map as BSS and the one-way model. The map implemented is the same only for the particular input state $\rho_{in}=\ket{0}\bra{0}$, with $\rho_{CTC}=\rho_{out}=\ket{+}\bra{+}$. For this input state Deutsch's CTC qubit is not entangled with the time-respecting qubits.

This suggests that the root of the problem with Deutsch's model is the incompleteness of the description of the CTC qubit. Deutsch's model prescribes that the CTC qubit be sent back in time as a mixed density matrix, which results in information loss about its prior interactions. This is naturally taken care of in BSS's model, as teleportation preserves the CTC qubit's entanglement with other systems, which was created via the unitary interaction $V$.


In the recent papers \cite{Ralph07, RalphM10}, Ralph and Myers have proposed an extension of Deutsch's model, in which one can heuristically describe the situation from the point of view of the  time-travelling qubit, which interacts with an infinite number of copies of itself.  By adding decoherence to this system, a unique solution is selected out of Deutsch's possibly many self-consistent solutions. This extension of Deutsch's model is not, however, sufficient to make it compatible with the predictions of the BSS model and one-way quantum computation. A simple way to see that is to note that according to BSS the CTC output is always in a pure state, whereas the multiple solutions proposed by Deutsch are typically mixed.

\section{Conclusion}

In summary, we have shown how CTCs appear in a natural way in the one-way model of measurement-based quantum computation. We studied a simple example of such CTCs using the Bennett/Schumacher/Svetlichny (BSS) CTC model, whose predictions agreed with those required by the one-way model. Going beyond the simple example we studied, we characterized a class of CTC circuits that admit deterministic BSS model simulations. The simulations can be found using stabilizer techniques associated with the one-way model.

We have also worked out the predictions of Deutsch's model for the same CTC example and found a general disagreement in comparison with what is expected from the BSS/one-way model. This incompatibility stems from Deutsch's incomplete description of the state being sent back in time, whose complete history of previous interactions is preserved by the teleportation step used in BSS's model.

\acknowledgments
We would like to acknowledge Joe Fitzsimons, Damian Markham and Daniel Jonathan for helpful discussions. We also would like to thank Vincent Danos for having shared his notes on local complementation. This work was partially funded by Brazilian agencies FAPERJ and CNPq.

\appendix*

\section{}
In this Appendix we work out an example of the stabilizer manipulations required to turn a (probabilistic) BSS CTC simulation circuit into a deterministic one. We will do this for the BSS circuit in Fig. \ref{fig bigbss}, highlighting some general features of the procedure.

Let us start by translating the circuit in Fig. \ref{fig bigbss} into a one-way pattern using the techniques described in \cite{BroadbentK09}. In comparison with usual quantum circuits, BSS circuits include two new elements: (1) introduction of auxiliary qubits prepared in state $\ket +$; (2) corresponding projections onto $\ket +$. The former is translated as command $N_i$ and the latter is translated as a deterministic measurement with anachronical correction $P^{\ket +}=M^0_i Z_i^{s_i}$. It is convenient to translate each $J_{\theta}$ gate on qubit $i$ as an equivalent command sequence involving an anachronical correction: $ M^{-\theta}_iZ^{s_i}_i {\wedge}Z_{ij} N_j$. In this way, and following the prescription of \cite{BroadbentK09}, each $J_{\theta}$ gate in the circuit requires an added ancilla.

Using the techniques described in \cite{BroadbentK09} we obtain the following command sequence, corresponding to the BSS circuit in Fig. \ref{fig bigbss}:
\begin{eqnarray}
&&M_1^{0}Z^{s_1}_1  M_5^{0}Z^{s_5}_5  M_3^{\theta}Z^{s_3}_3 \ket{G}, \label{eq seqt}\\ \nonumber \\ \nonumber
\ket{G}&\equiv& {\wedge}Z_{34} {\wedge}Z_{35} {\wedge}Z_{32} {\wedge}Z_{51} {\wedge}Z_{12} \\
&&N_1 N_2  N_4 N_5 \ket{\psi_{in}}_3, \label{eq seqt2}
\end{eqnarray}
where the qubits' sub-indices match those of the input qubits in the circuit of Fig. \ref{fig bigbss}, except qubit 5 which is a new qubit added in the pattern so as to implement the $J$ gate. Qubit 3 is our input state, and all the others are prepared in state $\ket{+}$ by the $N_j$ command. 

\begin{figure}
 \centering
 \includegraphics[scale=0.35]{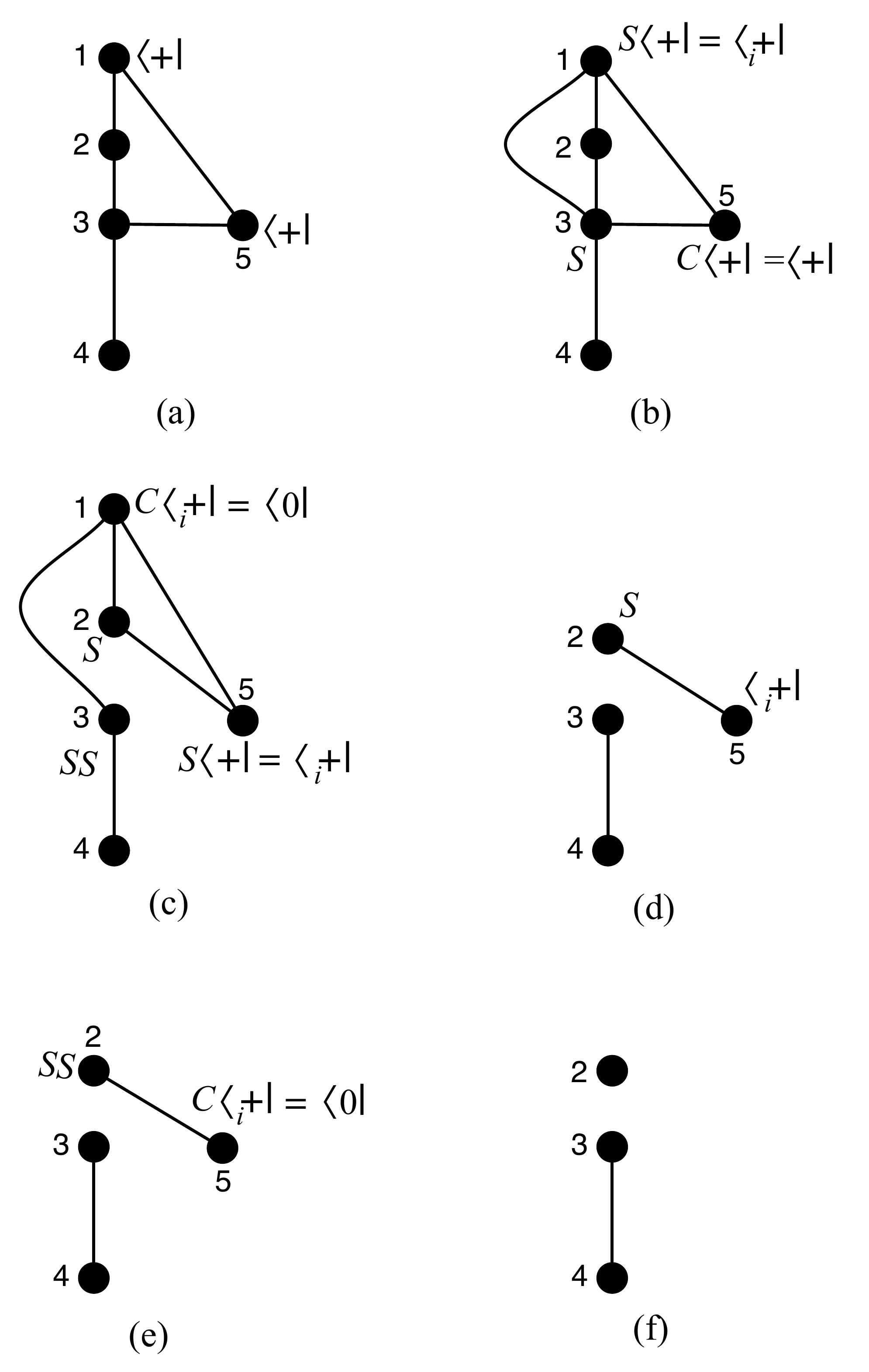}
 \caption{Stabilizer manipulations that simplify the one-way sequence (\ref{eq seqt}); $\bra{_i +}$ denotes a projection onto $\ket{0}+i\ket{1}$. a) Graph representing initial entanglement structure and operations. Qubit 3 is the input and qubits 1 and 5 are measured in the $X$ basis. b) Effect of local complementation (LC) on qubit 5; c) LC on qubit 1; d) Pauli $Z$ deletion of qubit 1; 3) LC on qubit 5; f) Pauli $Z$ deletion of qubit 5. The final pattern represents the one-way two-qubit implementation of the $J_{-\theta}$ gate, see Fig. \ref{fig jcc}.}
 \label{fig lcexample}
\end{figure}

The graph corresponding to state $\ket{G}$ in eq. (\ref{eq seqt2}) is shown in Fig. \ref{fig lcexample}-a. This one-way computation can be simplified using the so-called \textit{local complementation} rule introduced in \cite{vandenNestDM04}, which we now review.

\begin{figure}
 \includegraphics[scale=0.12]{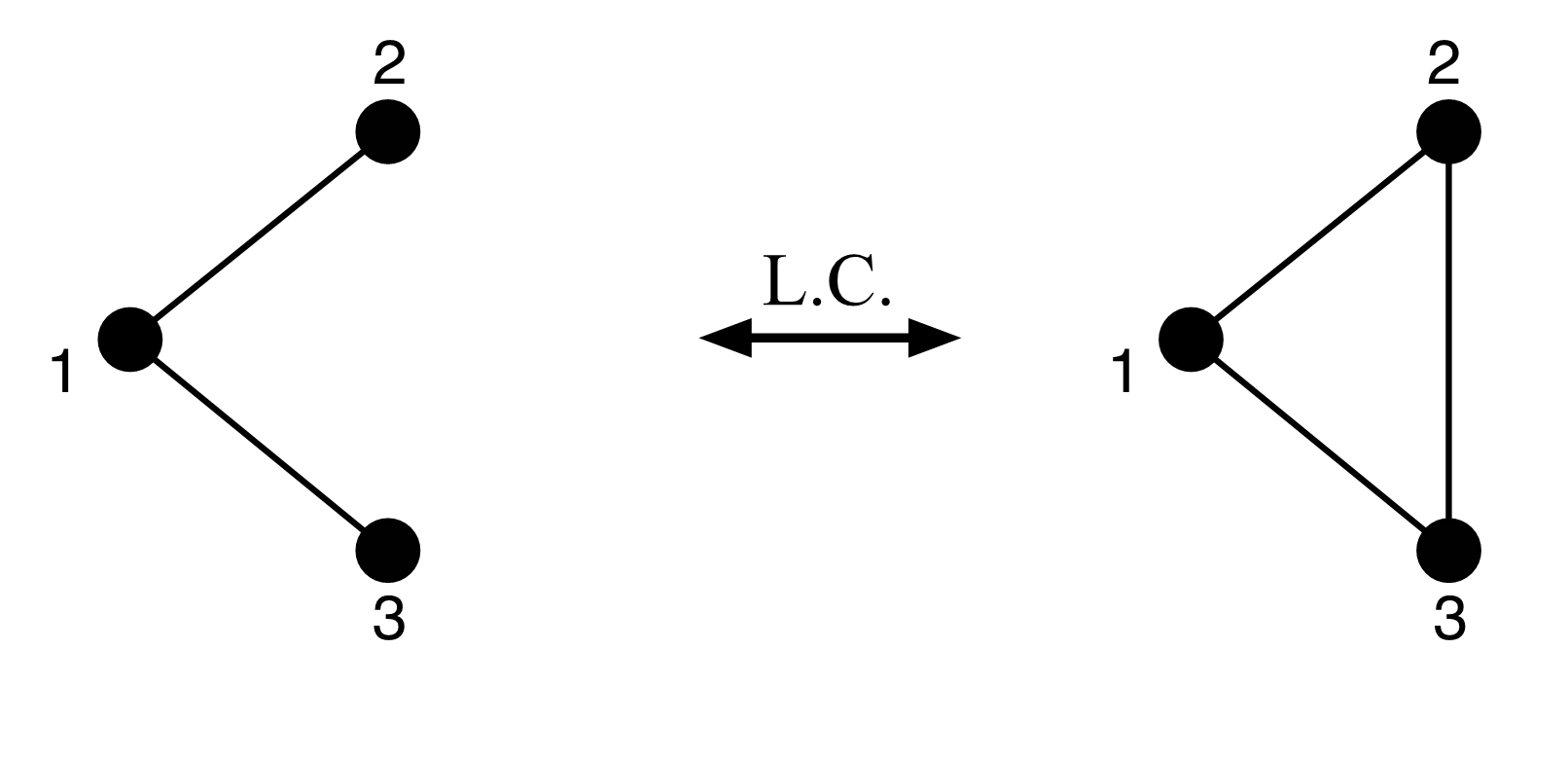}
 \caption{An analysis of the stabilizers in eqs. (\ref{kpsi}) and (\ref{kphi}) indicates that the graph on the right represents state $|\psi\rangle$ and the one on the left state $|\phi\rangle$ (see main text). The local complementation unitary $C_1 S^{\dagger}_2 S^{\dagger}_3$ changes the graph on the left into the graph on the right. Local complementations change the state without changing its entanglement structure or the one-way computations implementable by it.}
\label{fig lc}
\end{figure}

Local complementation is an operation that changes a state in a way that is most conveniently described by the change in its stabilizers. It corresponds to local unitaries applied on a chosen qubit and its neighbors in the graph, and has been shown to preserve the computation that can be performed using the state in the one-way model \cite{HeinDERvdNB06,vandenNestDM04, RaussendorfBB03}. First, let us recall the definition of the phase gate $S = |0\rangle\langle0| + i|1\rangle\langle1|$ and define the unitary $C = HSH = 1/\sqrt{2}(e^{\frac{i\pi}{4}}1 + e^{\frac{-i\pi}{4}}X)$, where $H$ is the one-qubit Hadamard gate. In the manipulations that follow we will use the following identities:
\begin{equation} \label{lc rules}
 SXS^{\dagger} = Y, CYC^{\dagger} = Z, CXC^{\dagger} = X.
\end{equation}
Let us now consider two different graph states $\ket{\psi}$ and $\ket{\phi}$, specified below by listing their stabilizers, respectively:
\begin{eqnarray}
\{f_1 = X_1 Z_2 Z_3, f_2 = Z_1 X_2 Z_3, f_3 = Z_1 Z_2 X_3\}, \label{kpsi} \\
\{f'_1 =X_1 Z_2 Z_3, f'_2 = Z_1 X_2, f'_3 = Z_1 X_3\}. \label{kphi}
\end{eqnarray}
It is easy to check that the unitary $ C_1 S^{\dagger}_2 S^{\dagger}_3$ applied on state $\ket{\phi}$ changes the stabilizers as follows:
\begin{equation}
\{g_1 = X_1 Z_2 Z_3, g_2 = Y_1 Y_2, g_3 = Y_1 Y_3\}.
\end{equation}
Note that the new stabilizers satisfy the relations $f_1 = g_1$, $f_2 = g_1 g_2$ and $f_3 = g_1 g_3$, hence $C_1 S^{\dagger}_2 S^{\dagger}_3\ket{\phi} = \ket{\psi}$. The application of local one-qubit unitaries such as $C$ and $S$ does not change a state's entanglement structure, but reveals that different one-way patterns can correspond to the same entanglement resource. In our example, we know states $\ket{\psi}$ and $\ket{\phi}$ satisfy the eigenvalue equations $f_i\ket{\psi} = \ket{\psi}$ and $g_i\ket{\phi} = \ket{\phi}$. It is straighforward to verify that these states can be represented by the two sequences of commands $\ket{\psi} = {\wedge}Z_{12} {\wedge}Z_{13} {\wedge}Z_{23} N_1^{0} N_2^{0} N_3^{0}$ and $\ket{\phi}= {\wedge}Z_{12} {\wedge}Z_{23} N_1^{0} N_2^{0} N_3^{0}$, where $N_i^{0}$ is the preparation of qubit $i$ in state $|+ \rangle$. Now we are able to illustrate graphically the local complementation rule we described above by constructing the graphs associated with the states $\ket{\psi}$ and $\ket{\phi}$ and relating them by the equality $ C_1 S^{\dagger}_2 S^{\dagger}_3 \ket{\phi} = \ket{\psi}$, as shown in Fig. \ref{fig lc}.

The local complementation operation on non-input qubit $i$ corresponds to applying the unitary $C_i\Pi_{j\in N(i)} S_{j}^{\dagger}$, with $N(i)$ being the set of vertices (qubits) which are neighbors of $i$ in the entanglement graph. In addition to local complementation, we can also delete a vertex from a graph by measuring it in the $Z$ basis \cite{RaussendorfBB03}. $Z$-deletion and local complementation together change a state without altering its entanglement structure, which preserves the computation being performed  \cite{Jozsa05, RaussendorfBB03, HeinDERvdNB06,vandenNestDM04}. In general, we will need to apply these operations in graphs with an arbitrary number of vertices and edges, choosing where the local complementation is needed and applying the rule accordingly.

In Fig. \ref{fig lcexample} we illlustrate a sequence of local complementations and $Z$-deletions that transforms the initial sequence (\ref{eq seqt}) (corresponding to the circuit of Fig. \ref{fig bigbss}) into a simpler sequence implementing the same unitary. Starting from figure \ref{fig lcexample}-a, we apply a sequence of local complementations operations in order to remove the $X$-measurements originated from the BSS protocol. The required operations are illustrated in Fig. \ref{fig lcexample}, so that in the end the $X$-measurements become $Z$-measurements, which can be removed. This procedure results in the following, equivalent command sequence:
\begin{equation}
M_3^{\theta}Z^{s_3}_3 {\wedge}Z_{34}  N_2  N_4 \ket{\psi_{in}}_3.
\end{equation}
Qubit $2$ does not participate in the computation, as it remains disentangled from the others. We see that elimination of the Pauli $X$ measurement results in exactly the same command sequence of eq. (\ref{eq stab1}). As we have seen in the main text, this sequence implements deterministically the unitary map $J_{-\theta}$.

\bibliographystyle{unsrt}

\end{document}